\documentclass[12pt]{iopart}
\usepackage[pdfpagemode=UseNone,pdfstartview=FitH,colorlinks=true]{hyperref}
\usepackage{enumitem}
\setlistdepth{9}
\usepackage{graphicx}
\usepackage{epstopdf}

\newlist{longenum}{enumerate}{5}
\setlist[longenum,1]{label=(\roman*)}
\setlist[longenum,2]{label=(\alph*)}
\setlist[longenum,3]{label=(\arabic*)}
\setlist[longenum,4]{label=(\roman*)}
\setlist[longenum,5]{label=(\alph*)}

\begin{document}
\title{Biophysics software for interdisciplinary education and research}

\author{J. M. Deutsch}
\address{Department of Physics, University of California, Santa Cruz CA 95064}
\ead{josh@ucsc.edu}

\begin{abstract}
Biophysics is a subject that is spread over many disciplines and
transcends the skills and knowledge of the individual student. This
makes it challenging both to teach and to learn. Educational materials
are described to aid in teaching undergraduates biophysics in an
interdisciplinary manner.  Projects have been devised on topics
that range from x-ray diffraction to the Hodgkin Huxley equations.
They are team-based and encourage collaboration.  The projects make
extensive use of software written in Python/Scipy which can be
modified to explore a large range of possible phenomena. The software
can also be used in lectures and in the teaching of more traditional
biophysics courses. 
\end{abstract}
\maketitle

\section{Introduction}
\label{sec:Introduction}

Many of the most active and productive areas in science are rapidly
becoming highly interdisciplinary, with teams of researchers from
diverse fields discovering new phenomena and creating new technology
that would be impossible within the confines of a single discipline.
The range of knowledge of an single individual necessary to create,
for example, rapid gene sequencing technology is beyond the capacity
of even the most gifted polymaths. Current research often involves
researchers taking place in separate areas It is now necessary to
have work on different aspects of a project, that require completely
different expertises, and then to combine these efforts to achieve
the final result.

Education has had a hard time keeping pace with the rapidly changing face
of technological and scientific advancement, with the majority of learning
taking place within a single field, and not addressing the important issues
involved with collaboration across very different fields. In this paper
I will describe the software and learning tools that have been developed to
create an interdisciplinary learning environment. It attempts to teach 
students from different disciplines how to collaborate in a way that mimics
the real challenges faced today in biophysics research.

The course  spanned one quarter, that is ten weeks, plus student
presentations, and involved approximately four hours of lectures and typically
three hours of discussion sections per week. The number of students in the
course ranged from 12 to 21.

Learning in this course focuses on several weekly projects that are studied
by interdisciplinary teams for students. Each team is normally composed
of three students from a wide range of backgrounds and majors. For example,
a physics, biochemistry, and biomolecular engineering major. The projects
use software written in SciPy~\cite{scipy,ScipyOliphant}, to serve multiple functions:
sometimes to perform simulations that act as artificial experimental input,
sometimes used to analyze data, and to gain theoretical and intuitive
understanding of biophysical processes in a way that requires active learning.
One student, typically, is responsible for a particular area, such as
computer coding, mathematical and physical analysis, or biological and
biochemical understanding. The problems are designed to promote interaction
between the different majors, to teach them about interdisciplinary collaboration
and allow them to exchange knowledge and approaches to scientific problems.

This course has now
been taught three times, and an evaluation of the efficacy of this approach
including the results of student assessments will be addressed in a future publication.
This work will focus specifically on the tools used. Not only is this useful
to other educators interested in using these tools, but also provides
a wealth of useful simulations that can be used on their own, and can be
useful in both teaching and in research. 

The material has been modified and enlarged as more experience on its efficacy
has been obtained. Because the author is a physicist and has had little
experience teaching the life sciences, an exceptionally talented
biochemistry major, Michelle V. Mai, who had previously taken this course,
was recruited to help in amending and designing questions that would be most beneficial 
to the biology and biochemistry undergraduates.
The course has been highly successful and has had a major impact on many of the
students who have taken it. It has often influenced their career choice, either
towards biophysics, or software development, and has obtained consistently high
evaluations from the students. The author believes that it is therefore
worthwhile disseminating the tools used in this course and describing how
it was taught.

\section{Uses of material}
\label{sec:Uses}

The main purpose of this paper is to introduce instructors to the 
biophysics material that I have developed.
The material develops a range of skills, in understanding
biological processes, physical laws, mathematical analysis, 
software development, and interdisciplinary collaboration. 
The material can be used in many different ways as I now
describe. 

The easiest use of this material is in lecture demonstrations.  A
variety of physical and biophysical processes can be demonstrated
and the relatively simple python source code allows for flexibility
in making changes, even during lecture. For this purpose, the
audience can be undergraduates at all levels, and graduate students.
It was found to be very useful to derive theoretical predictions and
then test these numerically in real time. Often it was found that
these simulations made theoretical discussion much more tangible
and led to a lot of questions and interaction during lectures.

It can be used to supplement already existing biophysics courses 
with material that students can use, either as homework, or as
special projects. The courses do not have to be interdisciplinary
in composition and can be composed of physics majors, but this will 
require modification of the problems. The material is often presented
with incomplete software, allowing students to fill in important
steps to complete the code. Instead the software can be supplied to students
in a completely working state. In that way, the simulations become
more like experiments, and potentially even lower division students
with a small amount of computer experience can benefit from some
of these projects.

Finally it can be used as I have, as the basis for a new biophysics course
at the upper division or graduate level. 

\section{Type of projects}
\label{sec:TypeOfProjects}

The projects spanned a wide range of problems in biophysics. 

Most projects were due in a week, but several were two week projects. 
Students would typically submit several projects every week. These
would often require the submission of a progress report at the end of a week
to ensure their timely completion. Some of the projects were tailored to
the research interests of students who were currently taking the course. For
example, the ``Distribution of ATP and mitochondria in axons" is a research
topic of current interest to William Saxton on the biology faculty at UCSC,
and a team of students had meetings with him to determine how to most usefully model 
this process.

Most of these have one or more SciPy programs associated with them, but some
were analytical, and one involved other software (Foldit). Aside from the
software component, there were physical and biological aspects that needed to
be addressed by the students. 

In order to be teachable in a space of ten weeks, there were three common
physical and mathematical themes that ran through the course that enabled the students to build 
on what they had learned: Fourier Transforms, diffusion, and molecular simulations.

The projects are described below in nearly chronological order, 
so as to explain in more detail how a fairly substantial territory was covered 
in a way that was accessible to the students with very diverse backgrounds.
\begin{figure}[htp]
\begin{center}
\includegraphics[width=\hsize]{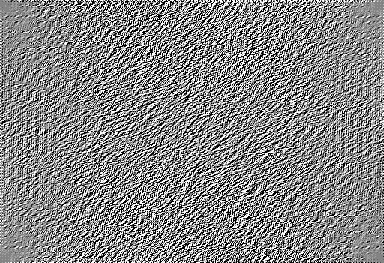}
\caption{
An image that needs to be decoded. The intensities are related to the Fourier
transform for the original image. This introduces the idea needed in x-ray diffraction
that inverting Fourier transform data can give you a lot of structural information.
}
\label{fig:fft_enhanced}
\end{center}
\end{figure}

\begin{figure}[htp]
\begin{center}
(a)\includegraphics[width=0.8\hsize]{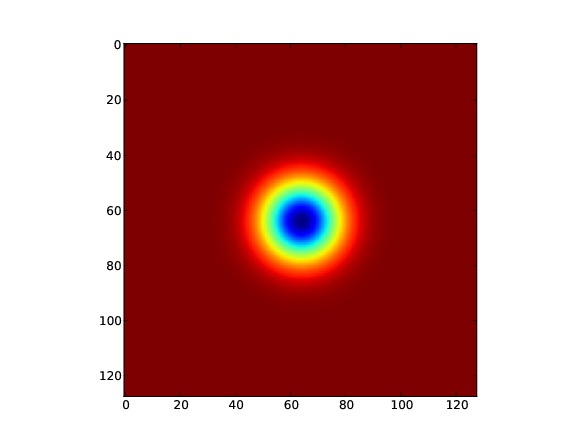}\\
(b)\includegraphics[width=0.8\hsize]{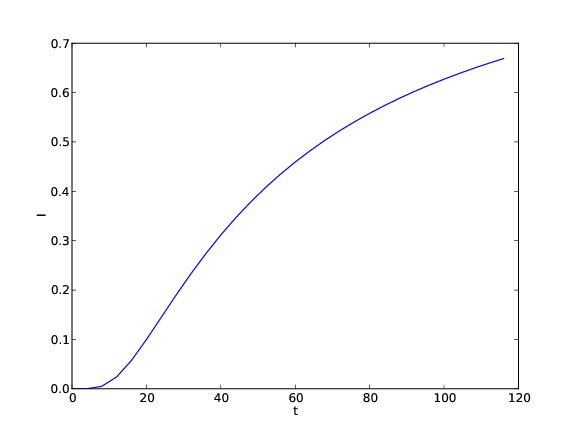}\\
\caption{
A simulation of FRAP. An area a circular area is photobleached and a
simulation of the evolution of density is displayed. (a) A snapshot of
the intensity during the simulation. (b) A plot of the intensity as a function
of time at the center of the circle.
}
\label{fig:FRAP}
\end{center}
\end{figure}

\begin{figure}[htp]
\begin{center}
\includegraphics[width=\hsize]{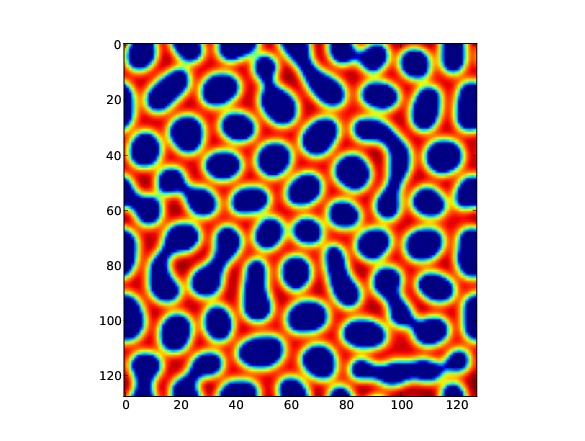}
\caption{
A simulation of FRAP. An area a circular area is photobleached and 
simulation of the evolution of density is displayed. (a) A snapshot of
the intensity during the simulation. (b) A plot of the intensity as a function
of time at the center of the circle.
}
\label{fig:patterns}
\end{center}
\end{figure}

\begin{figure}[htp]
\begin{center}
\includegraphics[width=\hsize]{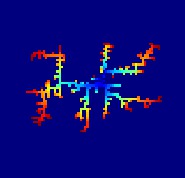}
\caption{
A simulation of Diffusion Limited Aggregation which is important in understanding
many physical processes including the morphology of bacterial colonies.
}
\label{fig:DLA}
\end{center}
\end{figure}

\begin{figure}[htp]
\begin{center}
\includegraphics[width=\hsize]{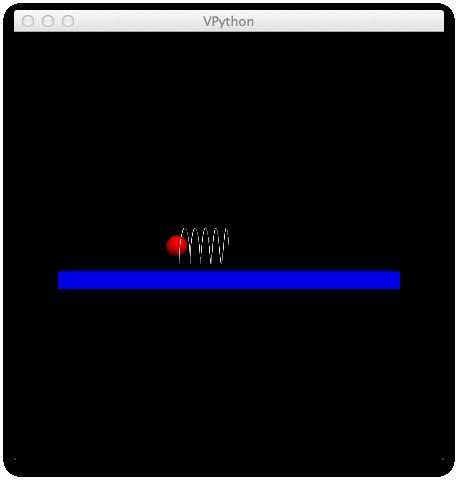}
\caption{
A snapshot of a simulation of the Langevin equation for a particle in a highly viscous
fluid coupled to a spring. This is the same mathematically as the model used to
understand noise in optical traps.
The graphics  was easily implemented using visual python~\cite{vpython}.
}
\label{fig:spring}
\end{center}
\end{figure}

\begin{figure}[htp]
\begin{center}
\includegraphics[width=\hsize]{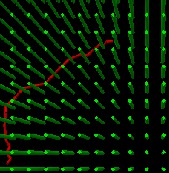}
\caption{
A snapshot of a simulation of electrophoresis of DNA going through an array
of nanopillars shown using visual python~\cite{vpython}.
}
\label{fig:elect}
\end{center}
\end{figure}

\begin{figure}[htp]
\begin{center}
\includegraphics[width=\hsize]{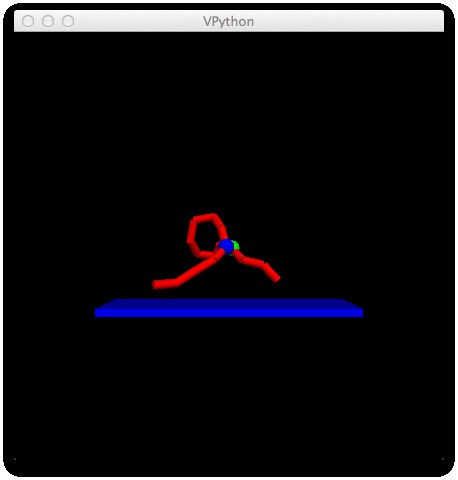}
\caption{
A snapshot of a simulation of a molecule with two metastable internal
states, to study Single Molecule Force Spectroscopy.
}
\label{fig:twostate}
\end{center}
\end{figure}

\subsection{Scope of projects}

The following goes through a synopsis of the material taught using
the materials discussed here. The plots and simulations shown are
a small subset of the ones available to the students. They are shown
as they would appear to them.

\begin{itemize}
\item [Week 1.]
The course started out with setting up SciPy on their individuals
computers.  They learned the basics of input/output by reading in
image files that contained coded images that could be decoded by
several operations including the two dimensional inverse Fourier
Transform. Fig. \ref{fig:fft_enhanced} shows the initial image that
needs to be decoded. The real part of the Fourier transform has been
processed in a way that allows for a reasonable quality when inverted, 
despite the small number of bits used to encode each point. When decoded
it leads to a picture that is recognizable to a life science student.
\item [Week 2.]
The next set of projects involves x-ray diffraction, and tomography,
both of which use Fourier Transforms. In two of the projects,
students were given fake computer generated data, (tomographic or
diffraction) and had to invert it to get the underlying structure.
The goal to find a structure mimics research more closely than book
problems that do not have such a tangible outcome.
\item [Week 3.]
Diffusion is introduced. The differential equations
describing this process are quite familiar to physics majors, but
the underlying connection with the physical process of random motion
is not usually explained. Computer projects introduced two methods
for understanding diffusion. One is from the differential equation
perspective, and the other using random walks. By going through
several of these projects, that contain a lot of visualization,
this connection becomes far clearer, and also introduces random
molecular motion to students. There is a plethora of biological
applications of diffusive processes and this also helps maintain
interest for students in the biological disciplines. One important 
technique, Fluorescence recovery after photobleaching~\cite{FRAP} (FRAP), is nicely
illustrated by computer simulation, (Fig. \ref{fig:FRAP}). A region
is photobleached and molecules diffuse resulting in a changing fluorescence
intensity. This project gets students to think about how a combination
of scaling and simulations can give quantitative predictions for
experiments. These techniques are rarely taught but can be appreciated,
without the need for sophisticated mathematics.
\item [Week 4.]
Most other problems could be leveraged from knowledge of these three
areas. In this week, various aspects of  morphogenesis
were discussed, and patterns such as those produced by reaction-diffusion
equations are now understandable by modifying the diffusion code
to include more than one kind of diffuser, and nonlinear interactions.
The model introduced by Turing giving rise to patterns on the skin of
animals can be modeled numerically (Fig. \ref{fig:patterns}) and give 
rise to a variety of different morphologies such as stripes and spots.
These can be understood by adjusting parameters in the models and the
boundary conditions. The most important lesson that comes out studying this
is how complexity comes out of simple physical processes.
Biology students were able to investigate to what extent these
principles are applicable to real problems. Other problems with
biological relevance, such as dendritic growth and Diffusion Limited
Aggregation~\cite{WittenDLA} (DLA) can now be taught, and simulations of these
can be modified by students to explore their behavior and relation
to biology. For example there has been much experimental work relating  
DLA (Fig. \ref{fig:DLA}) to the morphology of bacterial
colonies~\cite{MatsuyamaDLA,BenJacobDLA}. Students
not only learn how to write such simulations but by collaborating can critically 
assess evidence supporting this mechanism in experimental systems.
\item [Week 5.]
Instead of using differential equations to understand diffusion, a
more basic approach can be taken that uses the underlying random
motion. This starts the learner down the path of studying dynamics
of particles in liquids, taking into account their thermal motion.
To understand such problems, correlations and power spectra are
crucial, and this also relates back to Fourier Transforms. Particles
in optical trap provide an excellent illustration of these principles,
as well as being closely related to a very active area of biophysical
research. A simple example of this that can be easily simulated
is a particle tethered by a spring in a viscous liquid acted on by
random thermal motion (Fig. \ref{fig:spring}). This is perhaps the
most difficult part of the course to understand from a mathematical
perspective, and the projects employ simulations of these systems,
to obtain data as if they were experiments. By analyzing these
experiments by means of Fourier Transform and correlation
functions~\cite{Reif}, students appeared to grasp the underlying
physics quite well. This is also related to the more difficult
problem of bacterial chemotaxis, for example, how bacteria move to
regions of higher glucose concentration despite being too small to
be able to effectively detect chemical gradients. Many of the same
ideas can be applied to this interesting biological problem.

\begin{figure}[htp]
\begin{center}
\includegraphics[width=\hsize]{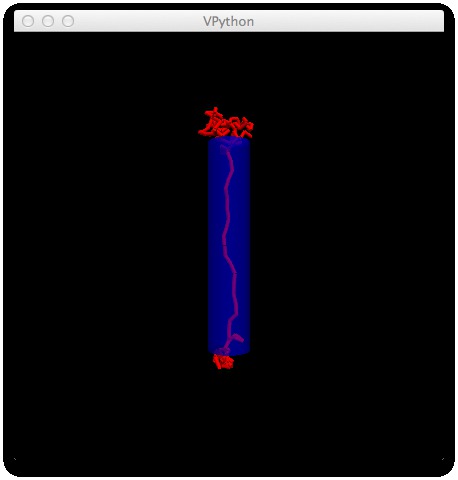}
\caption{
A snapshot of a simulation of a macromolecule translocating
through pore in a membrane. It is visualized using visual python~\cite{vpython}.
}
\label{fig:pore}
\end{center}
\end{figure}

\begin{figure}[htp]
\begin{center}
\includegraphics[width=\hsize]{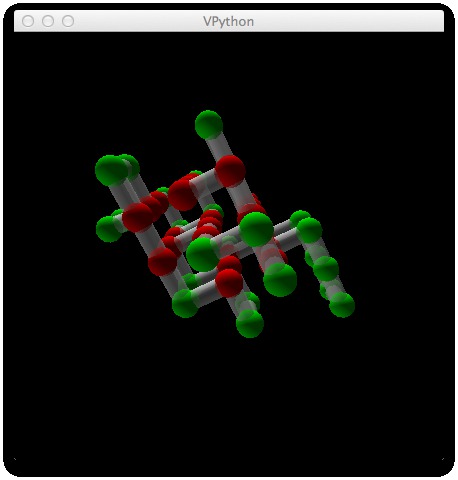}
\caption{
A snapshot of the folding of a protein on a lattice, using the HP
model, and Monte Carlo and simulated annealing. 
}
\label{fig:lattice_protein}
\end{center}
\end{figure}

\begin{figure}[htp]
\begin{center}
\includegraphics[width=\hsize]{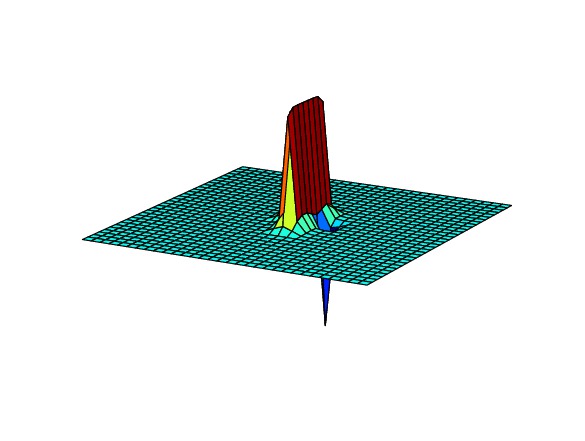}
\caption{
The potential around a collection of charges in two dimensions according to
the non-linear Poisson Boltzmann equation.
}
\label{fig:poissboltz}
\end{center}
\end{figure}

\item [Week 6.]
The above systems contain only a few degrees of freedom. 
The next logical step is to consider many
degrees of freedom, such as the dynamics of macromolecules. There
are a large number of biophysical systems that can be readily
investigated by simulations and are highly educational. The
electrophoresis of DNA can be described by simulating a 
string going through a network of rigid obstacles. The dynamics
show semi-periodic behavior~\cite{DeutschElect}, that can be easily seen by simulations
visualized in three dimensions using visual python~\cite{vpython}, see Fig.
\ref{fig:elect}.
This leads to an understanding of pulsed field electrophoresis,
which has been instrumental in separation of very long DNA that
cannot be separated by the constant field case. Using a modification
of this code, it is possible to understand the force versus
displacement in stretching a DNA chain which is accessible
experimentally using optical traps. More generally, recent experiments
stretching of macromolecules, such as RNA, that have complex internal
states, leads to a much increased understanding of their structure.
This technique, Single Molecule Force Spectroscopy~\cite{BlochSMFS}, is also possible
to understand directly by simulations (Fig. \ref{fig:twostate}. In this
case a molecule has two groups that have an attractive interaction so that when
it is pulled by its ends, it has metastable behavior giving rise to  hysteresis loops. 
Conversely, if one is given
experimental data on the stochastic transitions between multiple
internal states, analyzing the details of these transitions gives
much useful information about the nature of the molecule. This is
also examined in another project. Other projects have also been
developed, including the translocation of linear macromolecules
through a nanopore (Fig. \ref{fig:pore}), which is an active area of research at UC Santa
Cruz and elsewhere~\cite{AkesonPore}.

\begin{figure}[htp]
\begin{center}
(a)\includegraphics[width=\hsize]{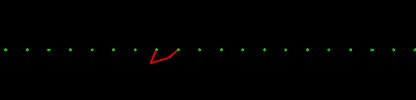}\\
(b)\includegraphics[width=\hsize]{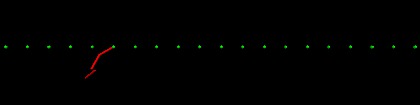}
\caption{
Two snapshots (a) and (b) of a minimal simulation of myosin V on actin
capturing the essential physics of motor motion. The motion
is similar to hand-over-hand motion used on ``monkey bars".
}
\label{fig:motor}
\end{center}
\end{figure}

\item [Week 7.]
One of the most widely studied class of macromolecules are proteins. Because
of their emphasis in biophysics, and their fascinating properties, a number of
projects were developed.
A number of important physical phenomena are
studied here, the coil-globule transition, the helix-coil transition, and
the determination of the folded state of a protein. This problem of ``protein
folding" is studied for simplified models such as the so-called ``HP" model
of Dill and Chan~\cite{HPDillChan}, see Fig. \ref{fig:lattice_protein}. This is a good model to simulate using
Monte Carlo methods, and leads to an understanding of the energy landscape
of this problem. The problem is still interesting with further simplification, studying it in two dimensions.
A powerful variations of Monte-Carlo, such as parallel tempering
is also studied. To speed up these simulations, C code is embedded using
``weave"~\cite{weave}, which introduces some of the more computer oriented
students to C programming. A phenomenal piece of software that folds proteins
called ``Foldit" which is freely available~\cite{foldit} is excellent for elucidating many
of the subtle interactions seen in proteins. It has been written in the
form of a computer game and has been very popular and players have produced some excellent
structures~\cite{folditWork}.

\item [Week 8.]
Some of the more subtle features of macromolecular interaction are electrostatic
and these can be approximately described in some cases by the non-linear Poisson Boltzmann
equation~\cite{PoissonBoltzmann}. Students learn how to solve this equation, whose solution closely
resembles earlier work they did on morphogenesis (Fig. \ref{fig:poissboltz}). There are
also two projects involving biological motors. The first is a minimal model of myosin V~\cite{DunnMyosinV}
that can be understood by a simulation incorporating chain stiffness, binding
and unbinding from a linear array of sites. Students can change parameters in
the model, such as bounding angles, and chain stiffness, to see what motors
work best. Two snapshots from such a simulation are given in  (Fig. \ref{fig:motor}).
Different teams design motors with different parameters and run a race to see which motor works best. Another
problem that is currently being studied at UC Santa Cruz~\cite{SerbusStreaming,DeutschStreaming} is that of cytoplasmic
streaming in drosophila oocytes. This can be modeled by a variation of the
DNA simulations done in week 6. This leads to an interesting set of behavior
such as a breaking of chiral symmetry, and collective organization of
microtubules to advect fluid chaotically through the cell. Fig. \ref{fig:egg}
shows a snapshot of a microtubule moving as a rotating helix due to the action of kinesin motors
walking upwards along its surface.

\begin{figure}[htp]
\begin{center}
\includegraphics[width=\hsize]{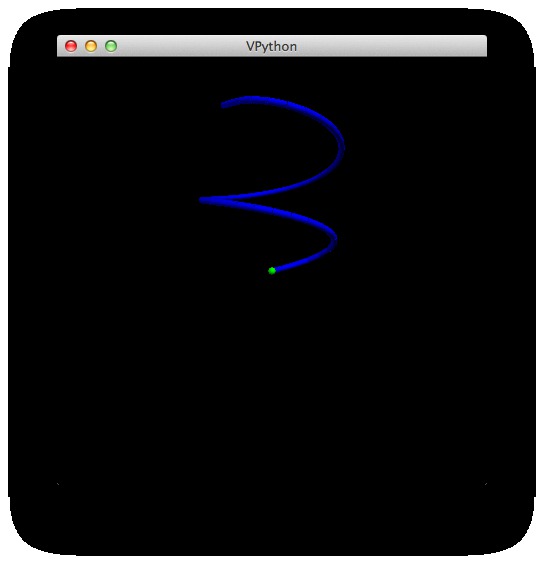}
\caption{
A snapshot of a microtubule moving inside a Drosophila oocyte. The
rotating helical motion is a consequence of the force generated
by kinesin motors walking along its backbone.
}
\label{fig:egg}
\end{center}
\end{figure}

\begin{figure}[htp]
\begin{center}
\includegraphics[width=\hsize]{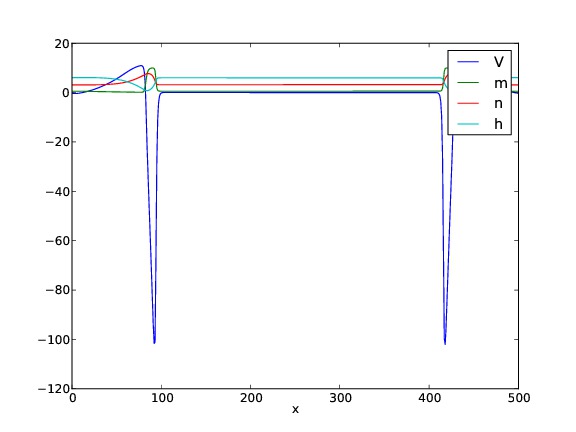}
\caption{
A snapshot of a simulation of action potentials travelling down an
axon as described by the Hodgkin Huxley equations.
}
\label{fig:hh_wave}
\end{center}
\end{figure}

\item [Week 9.]
The course ends with an introduction to biophysics of the neuron.
The Hodgkin Huxley equations~\cite{HodgkinHuxley} are modelled to explain how action potentials
are generated and propagate. The propagation of a spike is shown in Fig.
\ref{fig:hh_wave}. Complementary to this is the high level organization
of neural networks. Using code similar in spirit to that used earlier to study the Helix
Coil transition, the mechanism for associative memory proposed by Hopfield
is implemented. 
\end{itemize}

\subsection{List of projects}
Here is the list of projects offered in this course.


\begin{enumerate}
   \item Introduction 
   \begin{enumerate}
      \item Using SciPy and Inverse Fourier Transforming Images
   \end{enumerate}
   \item Imaging techniques
   \begin{enumerate}
          \item Reconstructing a 2D Structure From X-ray Data
	  \item Introduction to Tomography
	  \item Analytic methods in diffraction
	  \item Introduction to NMR (Two week project)
          \item Image processing
    \end{enumerate}
    \item Diffusion
    \begin{enumerate}
   	  \item {\bf 1 dimension}
          \begin{enumerate}
             \item  Binomial Distribution, Diffusion, Central Limit Theorem
             \item  Mean time to capture by diffusion in 1 dimension
             \item  Steady state solution of the diffusion eq in 1 dimension
             \item Comparing numerical and analytic solutions to the diffusion eq.
          \end{enumerate}
          \item {\bf 3 dimensions}
          \begin{enumerate}
               \item Probability of capture by a spherical absorber by diffusion
               \item Diffusion to a disk-like absorber
               \item Diffusion through many circular apertures in a planar barrier
          \end{enumerate}
          \item Using FRAP to determine a diffusion coefficient
	  \item Kirby Bauer Antibiotic Testing
	  \item Distribution of ATP and mitochondria in axons
    \end{enumerate}
    \item Morphogenesis
    \begin{enumerate}
       \item Reaction Diffusion and Biological Patterns
       \item Flow lines in Murray's model of pattern formation
       \item Morphogenesis by the mechanism of Diffusion Limited Aggregation
       \item Origins of dendritic growth
    \end{enumerate}
    \item Dynamics with thermal motion
    \begin{enumerate}
       \item Correlation functions and power spectra
       \item Brownian motion in an optical trap
       \item Brownian motion of a free particle
       \item Modeling bacterial chemotaxis
    \end{enumerate}
    \item Dynamics of Macromolecules 
    \begin{enumerate}
       \item Motion of DNA during gel electrophoresis
       \item Force on ends of DNA chain
       \item Translocation of a linear macromolecule through a nanopore
       \item Determining molecular states from force data
       \item Hysteresis in Single Molecule Force Spectroscopy
       \item Force on a freely jointed chain
    \end{enumerate}
    \item Understanding Protein interactions and folding
    \begin{enumerate}
       \item Foldit
       \item Coil Globule transition
       \item Helix Coil transition
       \item 2D HP model
       \item Folding time for the HP model 
       \item Using Parallel Tempering to fold a protein
       \item Effect of charges in solution
    \end{enumerate}
    \item Motor proteins
    \begin{enumerate}
       \item Myosin simulation race
       \item Cytoplasmic streaming in drosophila oocytes
    \end{enumerate}
    \item Membrane potentials and the Hodgkin Huxley Equations
    \item The Hopfield Model
\end{enumerate}

\section{Required student level of knowledge}
\label{sec:RequiredKnowledge}

At the beginning of the quarter, teams of three were formed, with the goal
to have enough diversity in knowledge in each team so as to effectively
tackle the projects assigned. 

\subsection{Software}

For each team, there was normally one student who had some degree
of exposure to programming. The physics majors are required to take
introductory programming and sometimes students from other majors,
particularly in the engineering disciplines were quite
knowledgeable.  The software was provided to the students, in variety
of states; sometimes as
fully functioning code, other times with a few lines that needed to be filled-in,
and rarely where more extensive modifications were needed to be made.

The software platform was chosen to follow a middle ground between two very
different pedagogical uses of computers in education. The first
scenario is exemplified by numerical methods courses. There, simulation
code is developed commonly in a highly efficient language, such as C,
C++, Fortran, or Java, to learn how to do state-of-the-art simulations
using a variety of numerical techniques, such as Monte Carlo, or
Molecular Dynamics simulations.  Such courses require students that typically
have a reasonably advanced knowledge of Applied Mathematics and the ability
to code at an intermediate level in one of the above languages. The
other scenario is to teach a course that already has a graphical
user interface (GUI) programmed in, and students navigate the
interface to make changes in parameters, or have a restricted
choice in how the software functions. Excellent biophysics software
of that type has been recently developed~\cite{TinkerXie} and
this approach has been pioneered more than two decades ago~\cite{Weiss}.

The software in this course is at neither of these two extremes.
It is recognized that the majority of physics and engineering majors have not had
a rigorous numerical methods course, but at the same time do typically
have a fair degree of programming proficiency. At the same time,
code even with the most flexible GUI interface, cannot come close to
the flexibility attainable using source code. It is very hard to have the
flexibility to allow students to answer the large number of questions
that come to mind, when exploring a new biophysical system. These
are the kind of questions that one would expect to answer when doing
research, and is indeed why SciPy has been so popular with the
research community. Scipy code is being extensively developed for use
in teaching by others as well~\cite{MyersSethna}. It allows fairly complex problems to be
addressed with a minimum of programming overhead, and also has
advanced graphics capabilities allowing data to be displayed in
a variety of different ways. Visual Python~\cite{vpython} was also
used. It displays excellent three dimensional visualization of
systems, such as macromolecules, with a minimum of extra code.

That being said, there was still a fair amount of learning that
was necessary to become proficient in the programming tools
that were used and the course was designed to make this possible
for the students to achieve. The learning of this skill was often
listed as a major benefit of having taken this course.

At the beginning of the course, the code was at a fairly basic
level, and eschewed the use of classes or other high level constructs
that might be unfamiliar to many of the students. As the quarter
progressed, the programs became more advanced and some 
C-code was integrated into the Python source through ``weave"~\cite{weave} to increase
its efficiency.

In general, the teams handled the software component of the projects
very well. In most cases students only needed to change parameters in the code,
and the students were able to follow the structure of
the code sufficiently well to be able to make these changes. In fact,
quite a few of the biology and biochemistry students were able to 
pick up enough programming to do this on their own. Some of them
ended up becoming quite proficient at coding.

There was quite a lot of choice students were given in projects that 
they would work on. Therefore the teams less interested in
software development would choose projects the did not require much
programming. On the other hand, some students were very interested
in this aspect, and occasionally produced very impressive modifications
or wrote their own code, for example, in Java. This flexibility in
coding emphasis led to more enthusiasm among the students,
who could often do work of an unexpectedly advanced nature.

The most difficult part of this course was the initial installation
of SciPy and Visual Python on students' own computers. These are both
complex pieces of software undergoing constant development. The operating
systems that worked best were Linux or Windows based. Mac computers
have been the most problematic. Some students were unable or unwilling
to upgrade their mac operating systems which made it much harder for
them to get a working version of SciPy. The latest releases of SciPy
from Enthought~\cite{enthought}, which has a free academic license,
worked well. However there are still major issues with Visual Python.
Some of these are problems with graphics cards, and with some
incompatibility of libraries between Visual Python and
the standard ones that are installed by Linux, and conflicts
between Enthought  and Visual Python on a mac.  Still, the vast majority 
of students found a way of getting a working system to enable them to modify 
and run the code for this course. The author has been able to get
the software to run on versions of all three operating systems, but because
of the nature of software, this is a constantly changing target.

\subsection{Knowledge of Biology and Biochemistry}

There was a significant fraction of students in this class that were not from
a physics background and primarily focused on biology or biochemistry. 

The course is focused on biology and therefore the problems that
need to be solved were at a level that requires a fair amount
of biological sophistication. An attempt was made, from the first assignment,
to ensure that these students' contribution was necessary
for a project's successful completion, even if a particular problem was mostly
one involving mathematics. For example, the first assignment required
a small modification in code in order to decode the Fourier Transform
of an image. The object was to determined what a set of images represents as
shown in Fig. \ref{fig:fft_enhanced}.
The last time the course was taught, the images were chosen to come
from biology or biochemistry, and it would have been quite hard for
physics majors to correctly identify them. This kind of interaction
between the different disciplines from the outset, was found to
greatly improve the cohesion of the teams and was later evident in
the increased degree of collaboration between members.

Aside from software or physics components, there were usually questions 
that required an upper division knowledge of biology or chemistry. These
questions were of a somewhat different character than you typically
find in biology courses. They often were more open-ended and allowed students
with different backgrounds to focus on different aspects of a problem.

There was an effort made to come up with problems that required a person
familiar with biological sciences to understand. 
Often it was necessary for teams to understand one or more biophysics articles in
order to undertake a project.  For example the
project to understand Fluorescence Recovery After Photobleaching (FRAP), which is
a common biophysical technique, is not taught to physics students.
References were given to papers that would be much more comprehensible
to biology or biochemistry students, who could then translate the
work into something a physics major could understand. Problems often
asked students to use their knowledge of biology to come up with situations
where a physical process, for example absorption of diffusers, would be
relevant, and to what extent such physical consideration would be
applicable. 

Because there was a fair amount of choice in projects, teams could 
choose problems more in line with their individual interests. There
were problems that were popular with neuroscience students for example,
such as studying MRI. Or there were specific problems, for example, the
translocation of a chain through a nanopore, or the Poisson Boltzmann equation,
that were of interest to students that had done laboratory research in that area.
These students could use their expertise to collaborate more effectively
with team members that were more software or mathematically oriented,
and this would be mutually beneficial to the whole team.

\section{Course materials}

The website for the course is
\href{http://physweb.ucsc.edu/drupal/courses/physics-180-spring-2012}{http://physweb.ucsc.edu/drupal/courses/physics-180-spring-2012}.
This contains the projects used in this course and other
information such as links to the material used in the lectures and
projects. The code provided to the students is, in some cases,
deliberately incomplete.
Instructors can email the author (josh@ucsc.edu) to obtain a full
set of python source code and data files.

\ack
\label{sec:Acknowledgements}
The author thanks  Dr. Barbara Goza and Hee-Sun Lee for useful
discussions. He especially would like to thank Michelle V. Mai for
her critical insights in making the material more relevant to life-science
majors.  This material is based upon work supported by the National Science Foundation
under Grant CCLI DUE-0942207.

{}

\end{document}